\documentclass[12pt]{article}
\usepackage{amsmath,amssymb}
\usepackage{graphicx}
\usepackage{amsmath}
\usepackage{hyperref}
\usepackage{float}
\usepackage{color, soul}
\sethlcolor{yellow}
\newcommand {\be}{\begin{equation}}
 \newcommand {\ee}{\end{equation}}
 \newcommand {\bea}{\begin{array}}
 
 \newcommand {\eea}{\end{array}}

\textwidth=6.5in \hoffset=-.75in \voffset=-1in \numberwithin{equation}{section}
\numberwithin{figure}{section}

\def\0{{(0)}}
\def\1{{(1)}}
\def\2{{(2)}}

\def\<{\langle }
\def\>{\rangle }
\def\[{\left[}
\def\]{\right]}

\begin{document}
\begin{titlepage}

\vskip1cm
\begin{center}
{~\\[140pt]{ \LARGE {\textsc{Holographic entanglement entropy and modular Hamiltonian in warped
CFT in the framework of GMMG model   }}}\\[-20pt]}
\vskip2cm

\end{center}
\begin{center}
{M. R. Setare \footnote{E-mail: rezakord@ipm.ir}\hspace{1mm} ,
M. Koohgard \footnote{E-mail: m.koohgard@modares.ac.ir}\hspace{1.5mm} \\
{\small {\em  {Department of Science,\\
 Campus of Bijar, University of Kurdistan, Bijar, Iran }}}}\\
\end{center}
\begin{abstract}
We study some aspects of a class of non-AdS holography where the 3d bulk gravity is given by Generalized Minimal Massive Gravity (GMMG). We consider the spacelike warped $AdS_3$ ($WAdS_3$) black hole solution of this model where the 2d dual boundary theory is the warped conformal field theory (WFCT).  The charge algebra of the isometries in the bulk and the charge algebra of the vacuum symmetries at the boundary are compatible and this is an evidence for the duality conjecture. Further evidence for this duality is the equality of entanglement entropy and modular Hamiltonian on both sides of the duality.
So we consider the modular Hamiltonian for the single interval at the boundary in associated to the modular flow generators of the vacuum. We calculate the gravitational charge in associated to the asymptotic Killing vectors that preserve the metric boundary conditions. Assuming the first law of the entanglement entropy to be true, we introduce the matching conditions between the variables in two side of the duality and we find equality of the modular Hamiltonian variations and the gravitational charge variations in two sides of the duality. According to the results of the present paper we can say with more sure that
the dual theory of the warped $AdS3$ black hole solution of GMMG is a
Warped CFT.      \end{abstract}
\vspace{1cm}

Keywords: Non-AdS holography, GMMG, Modular Hamiltonian, Modular flow generator, Generalized Rindler prescription

\end{titlepage}

\section{Introduction}	
One way to generalize the $AdS/CFT$ duality is to use the warped Anti-se Sitter (WAdS) spacetime in the gravity side of the theory. This is in the context of the so-called non-AdS holography. $WAdS_3$ are deformations of $AdS_3$  \cite{Beng} by very interesting applications \cite{Gui1,Hof1,Anni1}. One of the features of the warped AdS spacetime is that they exist in the near horizon of extremal Kerr black holes \cite{Gui1} and they realize black holes geometries \cite{Mou}. The $SL(2,R)\times U(1)$ group is the isometry group of these spaces.

The spacelike warped $AdS_3$ black hole is a solution of the generalized minimal massive gravity (GMMG) that has been introduced in the paper \cite{Setare2}. GMMG is a modification of the minimal massive gravity (MMG) \cite{MMG} that is an interesting theory in the context of $AdS/CFT$ correspondence where introduces a bulk theory with positive-energy propagating modes \cite{Setare2}. GMMG is a ghost free, pure gravity theory and it can avoid the so called bulk-boundary clash in a certain region of its parameter space. MMG and GMMG  are examples of  "third-way" consistent theories, i.e. their field equations do not come from an action, but the Bianchi identity is satisfied on-shell.\\

$WAdS_3/CFT_2$ duality has been investigated for the first time in \cite{Anni2} . The authors in \cite{Anni2} investigated the duality based on the entropy calculations utilizing the Cardy formula (for more details, one can refer to \cite{Son1,Det1} ). Investigation of the asymptotic behavior and the boundary conditions of the $WAdS_3$ has been done in \cite{Com}. The authors in \cite{GW}, utilizing the Lewkowycz-Maldacena procedure, find the holographic entanglement entropy in the non-(asymptotic) AdS spacetimes and they get the result explicitly for the $WAdS_3/CFT_2$ case.

The boundary conditions of the $WAdS_3/WCFT_2$ duality as another case of the non-AdS holographies in 3D spacetimes has been investigated in \cite{Com2}. The bulk geometry in the $WAdS_3/WCFT_2$ duality can be considered by the asymptotically warped $AdS_3$ spacetime that is dual to the warped conformal field theories (WCFTs) where the field theoretical futures of the duality has been investigated in \cite{Det1,Hofman1}. The WCFTs possess a global $SL(2,R)\times U(1)$ symmetry that leaves invariant the vacuum. They are 2d translation-invariant QFTs  with a chiral scaling symmetry that as a local symmetry have a Virasoro algebra plus a $U(1)$ Kac-Moody algbera. The $SL(2,R)\times U(1)$ symmetry is a subset of the global symmetries in CFTs given by $SL(2,R)\times SL(2,R)$. The local symmetries in CFTs are given by two copies of the Virasoro algebra \cite{Pol,Det1}. WCFTs possess an infinite dimensional group and the modular covariance that allow us to compute the entropy using a Cardy-like formula \cite{Card}. The entropy of a CFT using the Cardy formula is as follows \cite{Det1}
\begin{equation}\label{Card1}
  S_{CFT}= 2\pi\sqrt{\frac{c_R}{6}L_0}+ 2\pi\sqrt{\frac{c_L}{6}\bar{L}_0}
\end{equation}
where $L_0$ and $\bar{L}_0$ are the charges associated to the $SL(2,R)$ zero modes. $c_R$ and $c_R$  are the central charges of the algebra.\\

A universal result for the asymptotic entropy of a WCFT is as follows \cite{Det1}
\begin{equation}\label{Card2}
  S_{WCFT}=-\frac{4\pi i P_0P_0^{vac}}{k}+4\pi \sqrt{-(L_0^{vac}-\frac{(P_0^{vac})^2}{k})(L_0-\frac{P_0^2}{k})}
\end{equation}
where $L_0$ and $P_0$ are the charges associated to $SL(2,R)$ and $U(1)$ symmetries, respectively.  $k$ is the $U(1)$- level parameter. "vac" labels the vacuum expectation values of the charges. These formula (\ref{Card1}) and (\ref{Card2}) are analogous to each other but they are not equal.

The entanglement entropy and the modular Hamiltonian have important roles in establishment of $WAdS/WCFT$ duality and in verifying the first law of entanglement entropy. WCFTs have a non-relativistic nature. This property of the theory will be evident in the entanglement entropy of an interval in the vacuum, where we review the result of \cite{Castro} in the next section. This is based on the Rindler method that relies on the suitable coordinate maps \cite{Castro,Casini1}. A Rindler transformation is a warped conformal transformation that is satisfy some conditions. This transformation is invariant under a thermal identification. This is a symmetry of the vacuum and the vacuum is mapped to a state in the Rindler spacetime.

In parallel to the field theory calculations we have done in this paper, we investigate the holographic entanglement entropy and the gravitational charge. The holographic entanglement entropy of the $WAdS_3$ under the Dirichlet-Neumann boundary conditions has been computed in \cite{So1}, where the authors have used the Rindler method to find the result. In order to be able to reproduce the results of \cite{Castro} in the WCFT, the authors of \cite{So1} have considered an additional parameter to the Rindler transformations. They have found the result for the holographic entanglement entropy in dualities beyond the AdS/CFT. It can be seen the entanglement entropy in this case is different with the result of the Ryu-Takayanagi prescription \cite{Ta1,Ta2} which is based on a relation between the spacetime geometry and the entanglement entropy of the field theory side of the duality.

The general holographic picture of the $WAdS_3/WCFT_2$ duality has been analyzed in \cite{So2} by introducing the concept of the \emph{swing surfaces} that can be a modification to the Ryu-Takayanagi surfaces \cite{Ta1,Ta2} in the beyond AdS/CFT dualities. The swing surfaces can include two null geodesics that extend beyond the entangling interval at the boundary and a spacelike geodesic connecting the null geodesics.

Equality of the gravitational charge related to the isometries of the warped $AdS_3$ spacetime and the modular Hamiltonian of WCFT is an evidence to the holography conjecture. To compute the gravitational charge, we consider the metric as a spacelike warped $AdS_3$ black hole \cite{Tonni,Setare1}. The metric satisfies the asymptotically warped $AdS_3$ spacetime fall-off conditions \cite{Setare1,Henneaux}. Finding some dictionaries between the variables in two sides of the holography is an important step to establish the first law of entanglement entropy. This step is done by reading off the similar forms of the Fourier modes of the charges in two sides of the duality. So entanglement entropy can be considered as a test for $WAdS_{3}/WCFT$
conjecture, where we will study this test in the framework of GMMG model. In another term we investigate the holographic entanglement entropy and its first law in the context of $WAdS_{3}/WCFT$ correspondence for asymptotically warped $AdS_{3}$ solution of GMMG model.

This paper organized as follows. In section \ref{sec:2}, we review some properties of the WCFTs by defining the warped conformal transformations, Virasoro-Kac-Moody charge algebra and the modified form of the algebra. In the preceding subsection, we compute the modular flow generator $\zeta$ associated to a single interval in WCFTs on the plane. The modular flow can be defined as a linear combination of the vacuum symmetry generators. To find the universal form of the entanglement entropy, we review the calculation in \cite{Castro}. We have permission to use its result in our calculations to implement the holography. In section \ref{sec:3}, we show the first law of the entanglement entropy in the field theory side of the duality. In section \ref{sec:4}, we introduce the warped $AdS_3$ black hole as a GMMG solution and as an asymptotically warped $AdS_3$ spacetime. The asymptotic Killing vectors that preserving the boundary conditions will be introduced and the gravitational charge associated to this symmetry is computed. In section \ref{sec:5}, we find the first law of the entanglement entropy by introducing some holographic matching equalities between the variables on two sides of the $WAdS_3/CFT_2$ correspondence. In the last section, we provide a conclusion for the paper and we give a summary of our work on this class of non-AdS holography.
\section{Warped CFT}\label{sec:2}
We consider the warped conformal field theories (WCFTs) in two dimensions at the boundary of the gauge gravity duality. To this end, it is needed to have a good discussion on the general properties of these theories. To find an evidence for the holographic principles, we should have the entanglement entropy and the modular Hamiltonian, that we use the result of some earlier works. To provide everything we need, we give a review of the related works in \cite{Det1,Castro,Apolo} in this section.

\subsection{Some properties}\label{sec:2.1}
We consider a two dimensional theory defined on a plane with coordinates $(z,w)$. The spacetime symmetries are generated by the following transformations
\begin{equation}\label{transf}
  z'=f(z'),~~~w'=w+g(z'),
\end{equation}
where $f(z)$ and $g(z)$ are two arbitrary functions. The theories which are invariant under the transformations (\ref{transf}) are known as Warped Conformal Field Theories (WCFTs). The warped conformal transformation (\ref{transf}) is generated by two operators $T(z)$ and $P(z)$. The former is generator of infinitesimal coordinate transformation in \emph{z} and the latter is generator of \emph{z}-dependent infinitesimal coordinate translations in \emph{w}. $T(z)$ and $P(z)$ transform as follows
\begin{eqnarray} \label{tran.T}
% \nonumber % Remove numbering (before each equation)
  T'(z') &=& (\frac{\partial z}{\partial z'})^2\big[T(z)-\frac{c}{12}\{z',z\}\big]+\frac{\partial z}{\partial z'}\frac{\partial w}{\partial z'}P(z)-\frac{k}{4}(\frac{\partial w}{\partial z'})^2, \\
  P'(z') &=& (\frac{\partial z}{\partial z'})[P(z)+\frac{k}{2}\frac{\partial w}{\partial z'}], \label{tran.P}
\end{eqnarray}
where $\{.,.\}$ is the Schwarzian derivative as follows
\begin{equation}\label{Sch001}
  \{z',z\}=(\frac{\partial z'}{\partial z})^{-1}\frac{\partial^3z'}{\partial z^3}-\frac{3}{2}
  (\frac{\partial z'}{\partial z})^{-2}(\frac{\partial^2z'}{\partial z^2})^2.
\end{equation}

$c$ is the central charge and $k$ is the level of the Virasoro-Kac-Moody algebra \cite{Hofman1}. We can define the charges on the plane as follows
\begin{eqnarray}\label{W.Charges}
% \nonumber % Remove numbering (before each equation)
  L_n &=& -\frac{i}{2\pi}\int dz z^{n+1}T(z), \nonumber \\
  P_n &=& -\frac{1}{2\pi}\int dz z^{n}P(z).
\end{eqnarray}

These charges satisfy a Virasoro-Kac-Moody algebra with central charge $c$ and $U(1)$ level $k$ as follows
\begin{eqnarray}\label{VKM01}
% \nonumber % Remove numbering (before each equation)
  \[L_n,L_m\] &=& (n-m)L_{n+m}+\frac{c}{12}n(n^2-1)\delta_{n+m}, \nonumber \\
  \[P_n,P_m\] &=& \frac{k}{2}n\delta_{n+m}, \nonumber \\
  \[L_n,P_m\] &=& -m P_{n+m},
\end{eqnarray}
where $\delta_n$ is the Kronecker delta. The finite transformations (\ref{tran.T}) and (\ref{tran.P}) preserve the charge algebra (\ref{VKM01}) unchanged but this is not true for the vacuum expectation values of the zero-mode  charges. These transformations create changes on the vacuum expectation values that are called the spectral flow transformations. We are interested to find a mapping from the plane to the cylinder in coordinate $(z',w')$. The changes in the coordinates are as follows \cite{Castro}
\begin{equation}\label{cyl.01}
  z=e^{-iz'}, ~~~~w=w'+2\alpha z'.
\end{equation}

The Schwarzian derivative is obtained as follows
\begin{equation}\label{Sch.der}
  \{z',z\}=\frac{1}{2z^2}.
\end{equation}

Using eqs. (\ref{tran.T}), (\ref{tran.P}), (\ref{cyl.01}) and the Schwarzian derivative (\ref{Sch.der}), we find $T(z)$ and $P(z)$ transformations as follows
\begin{eqnarray} \label{P.cyl1}
% \nonumber % Remove numbering (before each equation)
  P^{cyl}(z') &=& i z P(z)-k\alpha \\
  T^{cyl}(z') &=& -z^2T(z)+\frac{c}{24}+2i\alpha z P(z)-k\alpha^2. \label{T.cyl1}
\end{eqnarray}

The modes on the cylinder can be defined as follows
\begin{eqnarray}\label{mod.def.cyl}
% \nonumber % Remove numbering (before each equation)
  L^{cyl}_n &=& -\frac{1}{2\pi}\int dz' e^{inz'}T^{cyl}(z'), \nonumber \\
  P^{cyl}_n &=& -\frac{1}{2\pi}\int dz' e^{inz'}P^{cyl}(z').
\end{eqnarray}

Substituting eqs. (\ref{P.cyl1}) and (\ref{T.cyl1}) in the charges on the cylinder, we find these charges in terms of the charges on the plane as follows
\begin{eqnarray}\label{rel.1}
% \nonumber % Remove numbering (before each equation)
  P^{cyl}_n &=& P_n+\alpha k \delta_{n,0} \\
  L^{cyl}_n &=& L_n+2\alpha P_n+(\alpha^2k-\frac{c}{24})\delta_{n,0}. \label{rel.2}
\end{eqnarray}

We can find the vacuum charges on the cylinder  using the above equations in terms of the central charge $c$ and the $U(1)$ level $k$, as follows
\begin{eqnarray}\label{Pvac.1}
% \nonumber % Remove numbering (before each equation)
  P_0^{vac}\equiv \langle P^{cyl}_0\rangle &=& \alpha k  \\
  L_0^{vac}\equiv \langle L^{cyl}_0\rangle &=& \alpha^2k-\frac{c}{24}. \label{Lvac.2}
\end{eqnarray}
where $\langle.\rangle$ is the notation for the expectation value on the vacuum states.

There is a modified algebra as follows \cite{Det1}
\begin{eqnarray}\label{VKM02}
% \nonumber % Remove numbering (before each equation)
  \[\tilde{L}_n,\tilde{L}_m\] &=& (n-m)\tilde{L}_{n+m}+\frac{c}{12}n(n^2-1)\delta_{n+m}, \nonumber \\
  \[\tilde{P}_n,\tilde{P}_m\] &=& 2n\tilde{P}_0\delta_{n+m}, \nonumber \\
  \[\tilde{L}_n,\tilde{P}_m\] &=& -m \tilde{P}_{n+m}+m\tilde{P}_0\delta_{n+m},
\end{eqnarray}
in which the $U(1)$ level $k$ is charge-dependent. We use this form of the algebra to analyze the duality in both sides of the gravity and the field theory. We could return to the original algebra (\ref{VKM01}) by the following re-definition of the charges  \cite{Det1}
\begin{eqnarray} \label{alg}
% \nonumber % Remove numbering (before each equation)
  \tilde{P}_n &=& \frac{2}{k}P_0P_n-\frac{1}{k}P_0^2\delta_n, \nonumber \\
  \tilde{L}_n &=& L_n-\frac{2}{k}P_0P_n+ \frac{1}{k}P_0^2\delta_n.
\end{eqnarray}

\subsection{Modular flow in WCFT  }\label{sec:2.1}
In this subsection we find the modular flow generator associated with a single interval in WCFT from the generalized Rindler
transformation \cite{Castro,Casini1,Jiang,Apolo}. To this end, we use a symmetry transformation of the vacuum
state to map an interval $\mathcal{I}$ to a Rindler spacetime. The modular flow generator is considered as a linear combination of the $SL(2,R)\times U(1)$ that leaves the vacuum invariant, as follows
\begin{equation}\label{Mod.fl01}
  \zeta = \Sigma a_i h_i
\end{equation}
where $a_i$ are some arbitrary constants. It is needed to satisfy the invariance of the causal domain under the modular flow transformations to find the coefficients $a_i$. We consider an interval $\mathcal{I}$ on the vacuum state
as follows
\begin{equation}\label{interval}
  \partial\mathcal{I}=\{(z_-,w_-),(z_+,w_+)\};~~~l_z=z_+-z_-, ~~l_w=w_+-w_-.
\end{equation}
where $(z,w)$ are the coordinates on the plane. This is the reference plane where the zero-mode charges vanish as
\begin{equation}\label{zero-mode01}
  P_0|0\rangle=L_0|0\rangle=0
\end{equation}
where $|0\rangle$ is the vacuum state. The symmetry generators $SL(2,R)\times U(1)$ that leave the causal domain $\mathcal{I}$ invariant are as follows
\begin{eqnarray}
% \nonumber % Remove numbering (before each equation)
  l_n &=& -z^{n+1}\partial_z;~~~~n=-1,0,+1 \nonumber \\
  \bar{l}_0 &=& -\partial_w,
\end{eqnarray}
where $\bar{l}_0$ is $U(1)$ generator and $l_n$ are the $SL(2,R)$ generators that satisfy the following algebra
\begin{eqnarray}
% \nonumber % Remove numbering (before each equation)
  \[l_-,l_+\] &=& 2l_0 \nonumber\\
  \[l_0,l_{\pm}\] &=& \pm l_{\pm}.
\end{eqnarray}

The modular flow generator as a linear combination of the vacuum symmetry generators can be writen as follows
\begin{equation}\label{zeta-1}
  \zeta=\zeta^z\partial_z+\zeta^w\partial_w=\Sigma_{j=-1}^1a_jl_j+\bar{a}_0\bar{l}_0.
\end{equation}

To find the coefficients $a_i$, it is needed the modular flow $\zeta$ vanishes at the boundaries of the causal domain. Satisfying this condition, we find the $a_i$ as follows
\begin{equation}\label{aj}
  (a_+,a_0,a_-)=a_+\big(1,-(z_++z_-),z_+z_-\big).
\end{equation}

To find the $a_+$, we have the conditions as follows \cite{Apolo}
\begin{eqnarray}\label{CDs}
% \nonumber % Remove numbering (before each equation)
  \partial_sz(s) &=& \zeta^z\nonumber \\
  z(s) &=& z(s+i)
\end{eqnarray}

This is because it is needed that $e^{i\zeta}$ maps a point in the causal domain back to itself. We use the first condition in (\ref{CDs}) as follows
\begin{equation}\label{dz/ds}
  \frac{dz}{ds}= -a_+(z^2-z_+z-z_-z+z_+z_-)
\end{equation}
where we have used (\ref{aj}). By integration of this equation and introducing the integration constants in $c_0$, we find the following result for $z(s)$
\begin{equation}\label{z01}
   z(s)=\frac{e^{a_+z_+s+c_0z_-}z_+-e^{a_+z_-s+c_0z}z_-}{e^{a_+z_+s+c_0z_-}-e^{a_+z_-s+c_0z}}
\end{equation}

Applying the 2nd condition in (\ref{CDs}), we find the following relation that is needed to satisfy this condition
\begin{equation}\label{rel.01}
  e^{a_+s(z_+-z_-)i}=1.
\end{equation}
Using the relation (\ref{rel.01}), we find the form of the $a_+$ as follows
\begin{equation}\label{a+}
  a_+=\frac{2\pi}{z_+-z_-}.
\end{equation}
Substituting the $a_+$ in (\ref{aj}), we find the final form of the coefficients as follows
\begin{equation}\label{aj02}
  (a_+,a_0,a_-)=\frac{2\pi}{z_+-z_-}\big(1,-(z_++z_-),z_+z_-\big).
\end{equation}

To find the $\bar{a}_0$, we pay attention to this point that this is the coefficient of $\bar{l}_0$ that creates translation along $w$ direction as follows
\begin{equation}\label{trans.w}
  w \to w-2\pi i \mu.
\end{equation}

To map a point in the causal domain back to itself, we shift the $w$ coordinate an amount $2\pi \mu$. Now we find the $\bar{a}_0$ as follows \cite{Apolo}
\begin{equation}\label{a.b0}
  \bar{a}_0=2\pi\mu.
\end{equation}

Substituting all the coefficients that we have found into (\ref{zeta-1}), we find the modular flow generator in the interval $\mathcal{I}$ as follows
\begin{eqnarray}\label{zeta-2}
% \nonumber % Remove numbering (before each equation)
  \zeta &=& 2\pi\mu\bar{l}_0+\frac{2\pi}{z_+-z_-}\big(l_1-(z_++z_-)l_0+z_+z_-l_{-1}\big) \nonumber \\
   &=& -2\pi\mu\partial_w-\frac{2\pi}{z_+-z_-}\big(z_+z_--(z_++z_-)z+z^2\big)\partial_z
\end{eqnarray}
where this is calculated in WCFT using the generalized Rindler method.

\subsection{Entanglement entropy in WCFT}\label{sec:2.2}
In this subsection we review the work of \cite{Castro} that is needed to implement the hologrpaphy in two sides of the bulk and the boundary. To calculate the entanglement entropy of a single interval, the background geometry is considered as a cylinder with coordinates $(t,x)$. The following identification defines the spatial circle
\begin{equation}\label{D01}
  (t,x)\sim(t+\bar{L},x-L),
\end{equation}
where the interval domain will be defined as follows
\begin{equation}\label{D02}
  \mathcal{D}: (t,x)\in\[(\frac{\bar{l}}{2},-\frac{l}{2}),(-\frac{\bar{l}}{2},\frac{l}{2})\]
\end{equation}

The entanglement entropy in $\mathcal{D}$ using the density matrix $\rho_{\mathcal{D}}$ is defined as follows
\begin{equation}\label{S.gen}
  S_{EE}=-Tr(\rho_{\mathcal{D}}\log\rho_{\mathcal{D}})
\end{equation}

Using a unitary transformation, the entanglement entropy is related to a thermal entropy \cite{Castro,Holzhey,Hartong}. For a warped system, the only allowed transformation is as follows
\begin{equation}\label{U.tran01}
  \frac{\tan \frac{\pi x}{L}}{\tan\frac{\pi l}{2L}}=\tanh\frac{\pi \tilde{x}}{L},~~~t+\frac{\bar{L}}{L}x=\tilde{t}+\frac{\bar{\kappa}}{\kappa}\tilde{x}
\end{equation}
where $\kappa$, $\bar{\kappa}$ are arbitrary scales and induces the following identification in $(\tilde{t},\tilde{x})$ coordinates
\begin{equation}\label{H01}
  \mathcal{H}: (\tilde{t},\tilde{x})\sim(\tilde{t}-i\bar{\kappa},\tilde{x}+i \kappa).
\end{equation}

The thermal density matrix of the domain $\mathcal{H}$ is related to the density  matrix of the interval $\mathcal{D}$ under the unitary transformation as follows
\begin{equation}\label{rhoD}
  \rho_{\mathcal{D}}=U\rho_{\mathcal{H}}U^{\dagger}
\end{equation}
where $\rho_{\mathcal{H}}=\exp(\bar{\kappa}P_0^{cyl}-\kappa L_0^{cyl})$. Using this relation between $\rho_{\mathcal{D}}$ and $\rho_{\mathcal{H}}$, the entanglement entropy can be defined as follows
\begin{equation}\label{S.gen02}
  S_{EE}=-Tr(\rho_{\mathcal{D}}\log\rho_{\mathcal{D}})=S_{thermal}(\mathcal{H}).
\end{equation}

Introducing a cutoff parameter $\epsilon$, the regulated interval $\mathcal{D}$ and its mapped domain $\mathcal{H}$ can be defined as follows \cite{Castro}
\begin{equation}\label{D03}
  \mathcal{D}: (t,x)\in\[(\frac{\bar{l}}{2}-\frac{\bar{l}}{l}\epsilon,-\frac{l}{2}+\epsilon),
  (-\frac{\bar{l}}{2}+\frac{\bar{l}}{l}\epsilon,\frac{l}{2}-\epsilon)\]
\end{equation}
and
\begin{equation}\label{H02}
  \mathcal{H}: (\tilde{t},\tilde{x})\in\[(\frac{\bar{\kappa}}{2\pi}\gamma-\frac{l}{2}\frac{\bar{L}}{L}+\frac{\bar{l}}{2},-\frac{\kappa}{2\pi}\gamma),
  (-\frac{\bar{\kappa}}{2\pi}\gamma+\frac{l}{2}\frac{\bar{L}}{L}-\frac{\bar{l}}{2},\frac{\kappa}{2\pi}\gamma)\],
\end{equation}
where
\begin{equation}\label{gam}
  \gamma=\log(\frac{L}{\pi\epsilon}\sin\frac{\pi l}{L})+O(\epsilon).
\end{equation}

To evaluate the thermal entropy, it is needed to find the partition function for $\mathcal{H}$
\begin{equation}\label{Z01}
  Z_{\bar{a}|a}(\bar{\theta}|\theta)
\end{equation}
where $(\bar{a},a)$ and $(\bar{\theta},\theta)$ are defined by the following indetifications
\begin{equation}\label{H03}
  (\tilde{t},\tilde{x})\sim(\tilde{t}+2\pi \bar{a},\tilde{x}-2\pi a)\sim(\tilde{t}+2\pi\bar{\theta},\tilde{x}-2\pi\theta)
\end{equation}
with
\begin{equation}\label{H04}
  2\pi a=\frac{\kappa}{\pi}\gamma,~2\pi\bar{a}=\frac{\bar{\kappa}}{\pi}\gamma-\frac{\bar{L}}{L}l+\bar{l},~~
  2\pi\theta=-i \kappa,~ 2\pi\bar{\theta}=-i\bar{\kappa}.
\end{equation}

The thermal entropy is defined as follows
\begin{equation}\label{SZ01}
  S_{\bar{a}|a}(\bar{\theta}|\theta)=(1-\theta\partial_{\theta}
  -\bar{\theta}\partial_{\bar{\theta}})\log Z_{\bar{a}|a}(\bar{\theta}|\theta).
\end{equation}

It could be seen the entropy is an observable that is independent of all observers. Changing the coordinates as follows
\begin{equation}\label{hatC}
  \hat{x}=\frac{\tilde{x}}{a},~~~\hat{t}=\tilde{t}+\frac{\bar{a}}{a}\tilde{x},
\end{equation}
the equality between the entropies is as follows \cite{Castro}
\begin{equation}\label{equl.S}
  S_{\bar{a}|a}(\bar{\theta}|\theta)=\hat{S}(\bar{\theta}-\frac{\theta}{a}\bar{a}|\frac{\theta}{a}).
\end{equation}

In the coordinates $(\hat{t},\hat{x})$, the partition function can be calculated as follows \cite{Castro}
\begin{equation}\label{Z02}
  \hat{Z}(\rho,\tau)=e^{i\pi\frac{k}{2}\frac{\rho^2}{\tau}}\hat{Z}(\frac{\rho}{\tau}|-\frac{1}{\tau})
  =e^{i\pi\frac{\kappa}{2}\frac{\rho^2}{\tau}}e^{2\pi i\frac{\rho}{\tau}P_0^{vac}+2\pi i\frac{1}{\tau}L_0^{vac} }+...
\end{equation}
where
\begin{equation}\label{ta.z}
  \tau=-i\frac{\pi}{\gamma},~~\rho=-\frac{i}{2\gamma}(\frac{\bar{L}}{L}l-\bar{l})
\end{equation}

In the 2nd equality in (\ref{Z02}), the vacuum dominates the sum. $\hat{Z}(\rho,\tau)$ is the partition  function with $(\bar{a},a)=(0,1)$ as a canonical partition function.

The thermal entropy has the following relation with the $\hat{Z}$ partition function,
\begin{equation}\label{SZ02}
  \hat{S}(\rho|\tau)=(1-\tau\partial_{\tau}
  -\rho\partial_{\rho})\log \hat{Z}(\rho|\tau).
\end{equation}

Using this relation, the thermal entropy can be defined as follows
\begin{equation}\label{S01}
  \hat{S}(\rho|\tau)=iP_0^{vac}l(\frac{\bar{L}}{L}-\frac{\bar{l}}{l})-4L_0^{vac}\log(\frac{L}{\pi \epsilon}\sin\frac{\pi l}{L})
\end{equation}

Substituting the thermal entropy in (\ref{S.gen02}), we find the entanglement entropy as follows
\begin{equation}\label{S02}
  S_{EE}=iP_0^{vac}l(\frac{\bar{L}}{L}-\frac{\bar{l}}{l})-4L_0^{vac}\log(\frac{L}{\pi \epsilon}\sin\frac{\pi l}{L})
\end{equation}
where $P_0^{vac}$ and $L_0^{vac}$ are the vacuum charge values on the cylinder. To evaluate the entanglement entropy of one segment in the WCFT at finite temperature, all we need to do is changing the map that we used before, as follows \cite{Castro}
\begin{equation}\label{U.tran02}
  \frac{\tanh \frac{\pi x}{L}}{\tanh\frac{\pi l}{2L}}=\tanh\frac{\pi \tilde{x}}{L},~~~t+\frac{\bar{\beta}}{\beta}x=\tilde{t}+\frac{\bar{\kappa}}{\kappa}\tilde{x}
\end{equation}

To find the entropy, we just to use the replacement $L\to i\beta$ and $\bar{L}\to i\bar{\beta}$. With this replacement in (\ref{S02}), we find the entanglement entropy as follows
\begin{equation}\label{S03}
  S_{EE}=i P_0^{vac}l(\frac{\bar{\beta}}{\beta}-\frac{\bar{l}}{l})-4L_0^{vac}\log(\frac{\beta}{\pi \epsilon}\sinh\frac{\pi l}{\beta}).
\end{equation}

\section{The first law of entanglement entropy in WCFT}\label{sec:3}
The modular Hamiltonian has the following relation with the partition function
\begin{equation}\label{Z03}
  Z=Tr\exp(-\mathcal{H}_{\zeta})=Tr\exp\big[2\pi i\rho P_0^{cyl}-2\pi i\tau L_0^{cyl}    \big]
\end{equation}
where we have used the following relation \cite{Castro}
\begin{equation}\label{Z.gen1}
  Z_{\bar{a}|a}(\bar{\tau}|\tau)=Tr_{\bar{a}|a}\bigg( \exp\big(2\pi i \bar{\tau}P_0^{cyl}-2\pi
      i \tau L_0^{cyl}\big)   \bigg)
\end{equation}

Using the relation (\ref{Z03}), we find the charge $\mathcal{H}_{\zeta}$ as follows
\begin{eqnarray}\label{Hm01}
% \nonumber % Remove numbering (before each equation)
  \mathcal{H}_{\zeta} &=& 2\pi i \rho P_0^{cyl}-2\pi i \tau L_0^{cyl}-\log Z \nonumber \\
   &=& -\frac{1}{2\pi} \int d\hat{x} \big\{2\pi i \rho\hat{P}(\hat{x})- 2\pi i \tau\hat{T}(\hat{x})\big\}-\log Z,
\end{eqnarray}
where in the 2nd line, we use the $L_n^{cyl}$ and $P_n^{cyl}$ definitions on the cylinder (\ref{mod.def.cyl}). $\mathcal{H}_{\zeta}$ is the charge related to the modular flow generator vector $\zeta$ (\ref{zeta-2}) that is given in the former subsection \ref{sec:2.2}. The modular Hamiltonian is given by the following relation \cite{Apolo}
\begin{equation}\label{H1,2}
  \mathcal{H}_{mod}=\mathcal{H}_{\zeta}+const.
\end{equation}
where the constant term makes the trace of the thermal density matrix $\rho_{\mathcal{D}}$ equal to one. To find the charge $\mathcal{H}_{\zeta}$ using (\ref{Z03}), we add $\log Z$ term to make the trace of the thermal density matrix equal to one. The modular Hamiltonian in terms of the $(t,x)$ coordinates variables can be obtained from (\ref{Hm01}). To this end, we should perform a series of warped conformal transformations $(\hat{t},\hat{x})~\to~(\bar{t},\bar{x})~\to~(t,x)$ using (\ref{tran.T}) and (\ref{tran.P}). The modular Hamiltonian $\mathcal{H}_{mod}$ can be find as follows \cite{Apolo}
\begin{eqnarray} \label{Hme.R01}
% \nonumber % Remove numbering (before each equation)
  \mathcal{H}_{mod} &=& \int_{x_-}^{x_+}dx\bigg\{2i\alpha P(x)+\frac{\beta}{2\pi}\big[
  \frac{\cosh\frac{\pi(2x-l)}{\beta}-\cosh\frac{\pi l}{\beta}}{\sinh\frac{\pi l}{\beta}}    \big]\big[
   T(x)-\frac{\bar{\beta}}{\beta}P(x)   \big]      \bigg\}\nonumber \\
   &+& ik\alpha\bar{l}-\big[ \frac{c}{12}-\frac{k}{8\pi^2}\bar{\beta}^2   \big]\big[\frac{\pi l}{\beta}
   \cosh\frac{\pi l}{\beta}-1   \big]
\end{eqnarray}
where $\alpha$ is defined as a characteristic parameter of the change of the coordinates (\ref{cyl.01}) between the plane and the cylinder. Consequently, the variation of the modular Hamiltonian can be find as follows
\begin{eqnarray} \label{Hme.R01}
% \nonumber % Remove numbering (before each equation)
  \delta\mathcal{H}_{mod} &=& \int_{x_-}^{x_+}dx\bigg\{2i\alpha \delta P(x)+\frac{\beta}{2\pi}\big[
  \frac{\cosh\frac{\pi(2x-l)}{\beta}-\cosh\frac{\pi l}{\beta}}{\sinh\frac{\pi l}{\beta}}    \big]\big[
   \delta T(x)-\frac{\bar{\beta}}{\beta}\delta P(x)   \big]      \nonumber \\
   &+& \frac{i k\alpha}{l}\delta\bar{l} \bigg\}
\end{eqnarray}

Using the $L_n^{cyl}$ and $P_n^{cyl}$ definitions on the cylinder (\ref{mod.def.cyl}), we could find the result as follows
\begin{eqnarray}\label{del.TP}
  \delta T(x) &=& -\delta L_0^{cyl},\nonumber\\
  \delta P(x) &=& -\delta P_0^{cyl}.
\end{eqnarray}

Substituting these relations into the modular Hamiltonian variation (\ref{Hme.R01}), we could find the variation as follows

\begin{eqnarray} \label{Hme.R02}
% \nonumber % Remove numbering (before each equation)
  \delta\mathcal{H}_{mod} &=& i \alpha k\delta\bar{l}-2i\alpha l\delta P_0^{cyl}-\frac{\beta^2}{4\pi^2}
  \big(2-\frac{2\pi l}{\beta}\coth\frac{\pi l}{\beta}   \big)\big(\delta L_0^{cyl}-\frac{\bar{\beta}}{\beta}\delta P_0^{cyl}     \big)      \nonumber \\
   &=& iP_0^{vac}\delta\bar{l}-2i\alpha l\delta P_0^{cyl}-\frac{\beta^2}{4\pi^2}
  \big(2-\frac{2\pi l}{\beta}\coth\frac{\pi l}{\beta}   \big)\big(\delta L_0^{cyl}-\frac{\bar{\beta}}{\beta}\delta P_0^{cyl}     \big)
\end{eqnarray}
where we have used the definition of $P_0^{vac}$ (\ref{Pvac.1}) in the 2nd line. The first law of entanglement entropy is a statement on the equality of the entanglement entropy variations
with the modular Hamiltonian variations as follows \cite{Apolo}
\begin{equation}\label{law01}
  \delta S_{EE}=\delta\langle \mathcal{H}_{mod}\rangle
\end{equation}
where we have the modular Hamiltonian variation $\delta\mathcal{H}_{mod}$ in (\ref{Hme.R02}). To show this law is satisfied in the warped CFTs, we need the variation of the entanglement entropy. The variation of the entanglement entropy (\ref{S03}) can be found as follows
\begin{equation}\label{delS01}
  \delta S_{EE}=iP_0^{vac}l\delta(\frac{\bar{\beta}}{\beta})-iP_0^{vac}\delta\bar{l}-4L_0^{vac}\big(1-\frac{\pi l}{\beta}\coth\frac{\pi l}{\beta}   \big)\frac{\delta \beta}{\beta}
\end{equation}

Substituting the following relations in (\ref{delS01}) \cite{Apolo}
\begin{eqnarray}\label{relR01}
  \frac{\delta\beta}{\beta^3} &=& -\frac{3}{c}\frac{1}{\pi^2}\delta\big(L_0^{cyl}-\frac{(P_0^{cyl})^2}{k}   \big),\nonumber\\
  \delta\big(\frac{\bar{\beta}}{\beta} \big) &=& \frac{2}{k}\delta P_0^{cyl},
\end{eqnarray}
we could find the variation of the entanglement entropy as follows
\begin{equation}\label{delS02}
  \delta S_{EE}=iP_0^{vac}\delta\bar{l}-2i\alpha l\delta P_0^{cyl}-
  \frac{\beta^2}{4\pi^2}\big(2-\frac{2\pi l}{\beta}\coth \frac{\pi l}{\beta} \big) \big(\delta L_0^{cyl}
  -\frac{\bar{\beta}}{\beta}\delta P_0^{cyl}     \big)
\end{equation}

Comparing the entanglement entropy variation (\ref{delS02}) and the variation of the modular Hamiltonian (\ref{Hme.R02}), we find these results are equal to each other. This is a proof of the first law of the entanglement entropy (\ref{law01}).

If we could find $\delta\bar{l}$ relation as follows
\begin{equation}\label{lb01}
 \delta\bar{l}=f(l,P_0^{vac}, P_0^{cyl})\delta P_0^{cyl},
\end{equation}
where $f(l,P_0^{vac}, P_0^{cyl})$ is a function of the $l$, $P_0^{vac}$ and $P_0^{cyl}$, we can introduce the following functions $K(x)$ and $Y(x)$ into (\ref{delS02})
 \begin{equation}\label{T01}
  K(x)=\big(\frac{\bar{\beta}}{\beta}   \big)\bigg[iP_0^{vac}f(l,P_0^{vac}, P_0^{cyl})\big(\frac{\beta}{\bar{\beta}}\big)-2i\alpha l\big(\frac{\beta}{\bar{\beta}}\big)+ \frac{\beta^2}{4\pi^2}\big(2-\frac{2\pi l}{\beta}\coth \frac{\pi l}{\beta} \big)\bigg]
\end{equation}
and
\begin{equation}\label{Y01}
  Y(x)=-\frac{\beta^2}{4\pi^2}\big(2-\frac{2\pi l}{\beta}\coth \frac{\pi l}{\beta} \big),
\end{equation}
then we have the entropy variation as follows
\begin{equation}\label{delS03}
  \delta S_{EE}=K(x)\delta P_0^{cyl}+Y(x)\delta L_0^{cyl}.
\end{equation}

We find the explicit form of the variation of $\bar{l}$ (\ref{lb01}) in the section \ref{sec:4}. Substituting the cylinder charges (\ref{mod.def.cyl}) into (\ref{delS03}), we find the following form of the entropy variation
\begin{equation}\label{delS04}
  \delta S_{EE}=-\frac{1}{2\pi}\int dx\bigg\{  K(x)\delta P+Y(x)\delta T\bigg\}.
\end{equation}

For later use, we find the variation of the entanglement entropy using the charges in the modified form (\ref{alg}). To this end, we find the relation between the variation of the charges $\tilde{P}_n$ and $\tilde{L}_n$ in the modified form and the variation of the charges $P_n$ and $L_n$ on the plane as follows
\begin{eqnarray}\label{def001}
% \nonumber % Remove numbering (before each equation)
  \delta\tilde{P}_0 &=& \frac{2P_0}{k}\delta P_0\nonumber \\
  \delta\tilde{L}_0+\delta\tilde{P}_0 &=& \delta L_0,
\end{eqnarray}
where we have used the relation (\ref{alg}) when $n=0$. The relation between the variations of the cylinder charges and the variations of the plane charges can be found as follows
\begin{eqnarray}\label{def002}
% \nonumber % Remove numbering (before each equation)
  \delta P_0^{cyl} &=& \delta P_0\nonumber \\
  \delta L_0^{cyl} &=& \delta L_0+2\alpha\delta P_0,
\end{eqnarray}
where we have used (\ref{rel.1}) and (\ref{rel.2}). Using (\ref{def001}) and (\ref{def002}), we could find the following relations between the charges on the cylinder and the charges in the modified form
\begin{eqnarray}\label{def003}
% \nonumber % Remove numbering (before each equation)
  \delta P_0^{cyl} &=& \frac{k}{2(P_0^{cyl}-\alpha k)}\delta \tilde{P}_0\nonumber \\
  \delta L_0^{cyl} &=& \delta\tilde{L}_0+\frac{P_0^{cyl}}{P_0^{cyl}-\alpha k}\delta \tilde{P}_0.
\end{eqnarray}

The relation between the variation of the zero-mode charges and the variation of the currents can be found as follows
\begin{eqnarray}\label{def004}
% \nonumber % Remove numbering (before each equation)
  \delta \tilde{P}_0 &=& -\delta\tilde{P},~~~~\delta P_0^{cyl}=-\delta P,\nonumber \\
  \delta\tilde{L}_0 &=& -\delta\tilde{T},~~~~\delta L_0^{cyl}=-\delta T,
\end{eqnarray}
where we have used (\ref{mod.def.cyl}) and the following relation for the modified charge.
\begin{eqnarray}\label{def005}
% \nonumber % Remove numbering (before each equation)
  \tilde{L}_0 &=& -\frac{1}{2\pi}\int dz' \tilde{T} (z'), \nonumber \\
  \tilde{P}_0 &=& -\frac{1}{2\pi}\int dz' \tilde{P}(z').
\end{eqnarray}

Substituting (\ref{def004}) into (\ref{def003}), we have the following result
\begin{eqnarray}\label{aalg}
% \nonumber % Remove numbering (before each equation)
  \delta P &=& \frac{k}{2(P_0^{cyl}-\alpha k)}\delta\tilde{P} \nonumber\\
  \delta T &=& \delta \tilde{T}+\frac{P_0^{cyl}}{P_0^{cyl}-\alpha k}\delta \tilde{P}.
\end{eqnarray}

Substituting (\ref{aalg}) into the entanglement entropy (\ref{delS04}), we find the modified form of the entropy as follows
\begin{equation}\label{SR04}
  \delta S_{EE}=-\frac{1}{2\pi}\int dx \big\{\big(\frac{K(x)k+2P_0^{cyl}Y(x)}{2(P_0^{cyl}-\alpha
  k)}\big)\delta \tilde{P}(x)+Y(x) \delta \tilde{T}(x)\big\}.
\end{equation}

 Now, we find the new form of the entanglement entropy as follows
\begin{equation}\label{SR05}
  \delta S_{EE}=-\frac{1}{2\pi}\int dx \big\{\tilde{K}(x)\delta \tilde{P}(x)+\tilde{Y}(x) \delta \tilde{T}(x)\big\}.
\end{equation}
where we have defined $\tilde{K}(x)$ and $\tilde{Y}(x)$ as follows
\begin{equation}\label{KR01}
  \tilde{K}(x)=\frac{K(x)k+2P_0^{cyl}Y(x)}{2(P_0^{cyl}-\alpha k)}
\end{equation}
and
\begin{equation}\label{YR01}
  \tilde{Y}(x)= Y(x).
\end{equation}

\section{Warped AdS black holes solution of GMMG  }\label{sec:4}
Generalized Minimal Massive Gravity (GMMG) was introduced in \cite{Setare2}, providing a new example of a
theory that avoids the bulk-boundary clash and therefore, as Minimal Massive Gravity (MMG)
\cite{MMG}, the theory possesses both, positive energy excitations around the maximally $AdS_3$ vacuum
as well as a positive central charge in the dual CFT. Since these theories avoid the bulk-boundary clash, they provide excellent areas to explore
the structure of asymptotically AdS solutions, asymptotic symmetries, their algebra and other
holographically inspired questions.
The warped $AdS_3$ black hole is a solution of the GMMG. The metric of this black hole is given by \cite{Tonni,Setare1}
\begin{equation}\label{Met01}
  ds^2=l^2\bigg(-N(r)^2dt'^2+\frac{dr^2}{4N(r)^2R(r)^2}+R(r)^2\big(d\phi+N^{\phi}(r)dt'\big)^2 \bigg)
\end{equation}
where $t'$, $r$ and $\phi$ are time-, radial- and angular-coordinates, respectively. We have used the $t'$-symbol instead of the $t$-symbol for the time coordinate, so as not to be confused with boundary calculations.  $l$ is the $AdS_3$ space radius. For the spacelike warped $AdS_3$ black hole we have the following definitions
\begin{eqnarray}\label{R01}
% \nonumber % Remove numbering (before each equation)
  R(r)^2 &=& \frac{1}{4}\alpha^2r\big[(1-\nu^2)r+\nu^2(r_++r_-)+2\nu\sqrt{r_+r_-}  \big] \\
  N(r)^2 &=& \alpha^2\nu^2\frac{(r-r_+)(r-r_-)}{4R(r)^2}, \label{N01} \\
  N^{\phi}(r) &=& |\alpha|\frac{r+\nu\sqrt{r_+r_-}}{2R(r)^2} , \label{Np01}
\end{eqnarray}
where $r_+$ and $r_-$ are the outer and inner horizon radiuses of the black hole, respectively. The metric (\ref{Met01}) has the symmetry group $SL(2, R) \times U(1)$ as the isometry group of the spacetime. This is the global part of the symmetry group of the Warped CFT. The appropriate boundary conditions to introduce asymptotically spacelike warped $AdS_3$ spacetime, is considered as follows \cite{Setare1,Henneaux}
\begin{eqnarray}\label{bC}
% \nonumber % Remove numbering (before each equation)
  g_{t't'} &=& l^2,~~~g_{tr}=\mathcal{O}(r^{-3}),~~~g_{r\phi}=\mathcal{O}(r^{-2}) ,\nonumber\\
  g_{t'\phi} &=& \frac{1}{2}l^2|\alpha|\big[r+A_{t\phi}(\phi)+\frac{1}{r}B_{t\phi}(\phi)  \big] ,\nonumber\\
  g_{rr} &=& \frac{l^2}{\alpha^2\nu^2}\big[\frac{1}{r^2}+\frac{1}{r^3}A_{rr}(\phi)+\frac{1}{r^4}B_{rr}(\phi)
  +\mathcal{O}(r^{-5})\big] ,\nonumber\\
  g_{\phi\phi} &=& \frac{1}{4}l^2\alpha\big[(1-\nu^2)r^2+rA_{\phi\phi}(\phi)+B_{\phi\phi}(\phi)\big]+\mathcal{O}(r^{-1})
\end{eqnarray}

These boundary conditions are in consistent with the metric (\ref{Met01}) by the $r$-dependent functions defined in (\ref{R01}), (\ref{N01}) and (\ref{Np01}). The warped $AdS_3$ black hole is an asymptotically spacelike warped $AdS_3$ spacetime. The asymptotic Killing vectors that generate the fluctuations preservintg the boundary conditions (\ref{bC}) are as follows \cite{Setare1}
\begin{eqnarray}\label{K.vec}
% \nonumber % Remove numbering (before each equation)
  \xi^{t'}(\mathcal{K},\mathcal{Y}) &=& \mathcal{K}(\phi)-\frac{2\partial^2_{\phi}\mathcal{Y}(\phi)}{|\alpha|^3\nu^4r}+\mathcal{O}(r^{-2}) , \nonumber\\
  \xi^r(\mathcal{K},\mathcal{Y}) &=& -r\partial_{\phi}\mathcal{Y}(\phi)+\mathcal{O}(r^{-2}) ,\nonumber\\
  \xi^{\phi}(\mathcal{K},\mathcal{Y}) &=& \mathcal{Y}(\phi)+\frac{2\partial^2_{\phi}\mathcal{Y}(\phi)}{\alpha^4\nu^4r^2}+\mathcal{O}(r^{-3}) ,
\end{eqnarray}
where $\mathcal{K}(\phi)$ and $\mathcal{Y}(\phi)$ are two arbitrary periodic functions. The corresponding conserved charge to the asymptotic Killing vectors (\ref{K.vec}) is given in \cite{Setare1} as follows
\begin{equation}\label{Q01}
  \mathcal{Q}(\mathcal{K},\mathcal{Y})=\mathcal{P}(\mathcal{K})+\mathcal{L}(\mathcal{Y}),
\end{equation}
where
\begin{equation}\label{gP01}
  \mathcal{P}(\mathcal{K})=-\frac{|\alpha|}{96\pi}c_U\int_{0}^{2\pi}\mathcal{K}(\phi)\big[A_{rr}(\phi)+2A_{t\phi}(\phi)\big]d\phi,
\end{equation}
and
\begin{equation}\label{gL01}
  \mathcal{L}(\mathcal{Y})=\frac{\alpha^4\nu^4}{768\pi}c_V\int_{0}^{2\pi}\mathcal{Y}(\phi)\big[-3A_{rr}(\phi)^2+4B_{rr}(\phi)+
  16B_{t\phi}(\phi) \big]d\phi,
\end{equation}
where
\begin{eqnarray}\label{cU}
  c_U &=& \frac{3l|\alpha|\nu^2}{G}\bigg\{\sigma+\frac{\omega}{\mu}\big(H_1+l^2H_2\big)+\frac{1}{m^2}
  \big(F_1+l^2F_2\big)-\frac{|\alpha|}{2\mu l}\bigg\}, \\
  c_V &=& \frac{3l}{|\alpha|\nu^2G}\bigg\{\sigma+\frac{\omega}{\mu}\big(H_1+l^2H_2\big)+\frac{1}{m^2}
  \big(F_1+l^2F_2\big)-\frac{|\alpha|}{2\mu l}\big(1-2\nu^2\big)\bigg\}. \label{cV}
\end{eqnarray}

$H_1$, $H_2$, $F_1$ and $F_2$ are constant parameters. $\sigma$ is a sign and $m$ is the mass parameter of New Massive Gravity (NMG) term \cite{NMG} in the Lagrangian of GMMG model. $\omega$ is a dimensionless parameter. Introducing $M(\phi)$ and $J(\phi)$ as follows
\begin{eqnarray}\label{M,J}
% \nonumber % Remove numbering (before each equation)
  M(\phi) &=& A_{rr}(\phi)+2A_{t\phi}(\phi) ,\nonumber\\
  J(\phi) &=& -3A_{rr}(\phi)^2+4B_{rr}(\phi)+16B_{t\phi}(\phi)  ,
\end{eqnarray}
we could change the form of $\mathcal{P}(\mathcal{K})$ and $\mathcal{L}(\mathcal{Y})$ in (\ref{gP01}) and (\ref{gL01}) , respectively as follows
\begin{equation}\label{gP02}
  \mathcal{P}(\mathcal{K})=-\frac{|\alpha|}{96\pi}c_U\int_{0}^{2\pi}\mathcal{K}(\phi)M(\phi)d\phi,
\end{equation}
and
\begin{equation}\label{gL02}
  \mathcal{L}(\mathcal{Y})=\frac{\alpha^4\nu^4}{768\pi}c_V\int_{0}^{2\pi}\mathcal{Y}(\phi)J(\phi)d\phi,
\end{equation}

In the case of the warped black hole (\ref{Met01}), we have
\begin{eqnarray}\label{A,B}
% \nonumber % Remove numbering (before each equation)
  A_{rr} &=& r_++r_-,~~~A_{t\phi}=\nu\sqrt{r_+r_-}, \nonumber \\
  B_{rr} &=& r_+^2+r_-^2+r_+r_-,~~~B_{t\phi}=0.
\end{eqnarray}

We define the Fourier modes as follows
\begin{eqnarray}\label{Frr01}
% \nonumber % Remove numbering (before each equation)
  \mathcal{P}_m &=& \mathcal{Q}(e^{im\phi},0)=\mathcal{P}(e^{im\phi}),\nonumber\\
  \mathcal{L}_m &=& \mathcal{Q}(0,e^{im\phi})=\mathcal{L}(e^{im\phi}) .
\end{eqnarray}

Using this definition, the charges can be found as follows
\begin{equation}\label{gP03}
  \mathcal{P}_m(\mathcal{K})=-\frac{|\alpha|}{96\pi}c_U\int_{0}^{2\pi}\mathcal{K}(\phi)M(\phi)e^{im\phi}d\phi,
\end{equation}
and
\begin{equation}\label{gL03}
  \mathcal{L}_m(\mathcal{Y})=\frac{\alpha^4\nu^4}{768\pi}c_V\int_{0}^{2\pi}\mathcal{Y}(\phi)J(\phi)e^{im\phi}d\phi.
\end{equation}

 The charge algebra can be written as follows \cite{Setare1}
 \begin{eqnarray}\label{VKM03}
% \nonumber % Remove numbering (before each equation)
  \[\mathcal{L}_n,\mathcal{L}_m\] &=& (n-m)\mathcal{L}_{n+m}+\frac{c_V}{12}n(n^2-1)\delta_{n+m}, \nonumber \\
  \[\mathcal{P}_n,\mathcal{P}_m\] &=& -\frac{c_U}{12}\mathcal{P}_0\delta_{n+m}, \nonumber \\
  \[\mathcal{L}_n,\mathcal{P}_m\] &=& -m \mathcal{P}_{n+m}+\frac{m}{2}\mathcal{P}_0\delta_{n+m}.
\end{eqnarray}

This charge algebra is the same as the modified algebra (\ref{VKM02}) that we have used in the field theory side provided that we introduce the following state-dependent map between the bulk $(\phi,t')$ coordinates and the boundary $(x,t)$ coordinates \cite{So1,Apolo}
\begin{equation}\label{lb02}
  \phi=x,~~~~t'=x+\frac{k t}{2\mathcal{P}_0},
\end{equation}
as a result, a mapping needs to be exist between the gravitational charges on the bulk and the warped conformal charges on the boundary, that we find the mapping in the next section.  This correspondence between the charge algebras on the bulk and the boundary is an evidence for the $WAdS_3/WCFT_2$ duality. Using (\ref{gP01}) and (\ref{gL01}) and from (\ref{A,B}), we can find the eigenvalues of the $\mathcal{P}_m$ and $\mathcal{L}_m$ as follows
\begin{eqnarray}\label{gP04}
% \nonumber % Remove numbering (before each equation)
  p_m &=& -\frac{|\alpha|c_U}{48}(r_++r_-+2\nu\sqrt{r_+r_-})\delta_m ,\\
  l_m &=& \frac{\alpha^4\nu^4c_V}{384}(r_+-r_-)^2\delta_m . \label{gL04}
\end{eqnarray}

\section{The first law of entanglement entropy in $WAdS_3$  }\label{sec:5}
We can write the gravitational charge of the GMMG as follows\cite{Setare1}
\begin{eqnarray}\label{Q02}
  Q(\mathcal{K},\mathcal{Y}) &=& -\frac{|\alpha|}{96\pi}c_U\int_{0}^{2\pi}\mathcal{K}(\phi)M(\phi)d\phi \nonumber \\
   &+& \frac{\alpha^4\nu^4}{768\pi}c_V\int_{0}^{2\pi}\mathcal{Y}(\phi)J(\phi)d\phi
\end{eqnarray}
where we have used (\ref{Q01}),(\ref{gP02}) and (\ref{gL02}) and as we have mentioned previously $\mathcal{K}(\phi)$ and $\mathcal{Y}(\phi)$ are two arbitrary periodic functions. The $M(\phi)$ and $J(\phi)$ function have been specified in (\ref{M,J}). The fourier modes of the gravitational charge can be found as follows
\begin{eqnarray}\label{Q03}
  Q_m(\mathcal{K},\mathcal{Y}) &=& -\frac{|\alpha|}{96\pi}c_U\int_{0}^{2\pi}\mathcal{K}(\phi)M(\phi)e^{i m \phi}d\phi \nonumber \\
   &+& \frac{\alpha^4\nu^4}{768\pi}c_V\int_{0}^{2\pi}\mathcal{Y}(\phi)J(\phi)e^{i m \phi}d\phi
\end{eqnarray}
The gravitational charge variations of (\ref{Q02}) can be found as follows
\begin{eqnarray}\label{Q04}
  \delta Q(\mathcal{K},\mathcal{Y}) &=& -\frac{|\alpha|}{96\pi}c_U\int_{0}^{2\pi}\mathcal{K}(\phi)\delta M(\phi)d\phi \nonumber \\
   &+& \frac{\alpha^4\nu^4}{768\pi}c_V\int_{0}^{2\pi}\mathcal{Y}(\phi)\delta J(\phi)d\phi
\end{eqnarray}
where we have considered the variations of lengths of bulk coordinates  as follows
\begin{equation}\label{lb03}
  \delta l_{\phi}=0,~~~~\delta l_{t'}=0.
\end{equation}

Utilizing the relation between the bulk and the boundary coordinates in (\ref{lb02}), we could find the following dictionary between the bulk and the boundary lengths of the coordinates
\begin{eqnarray}\label{lb2-1}
% \nonumber % Remove numbering (before each equation)
  l_{\phi} &=& l, \nonumber\\
  l_{t'} &=& l+\frac{k}{2\mathcal{P}_0}\bar{l}.
\end{eqnarray}

We could find the variations of the bulk and the boundary lengths, utilizing (\ref{lb2-1}), as follows
\begin{eqnarray}\label{lb2-2}
% \nonumber % Remove numbering (before each equation)
  \delta l_{\phi} &=& \delta l=0, \nonumber\\
  0 &=& \frac{k}{2\mathcal{P}_0}\delta \bar{l}-\frac{k\bar{l}}{2 \mathcal{P}_0^2}\delta \mathcal{P}_0.
\end{eqnarray}
where we have used (\ref{lb03}) into the above relations. Substituting the following relation into (\ref{lb2-2})
\begin{equation}\label{lb2-3}
  \bar{l}=\frac{2\mathcal{P}_0}{k}(l_{t'}-l),
\end{equation}
where we have obtained from (\ref{lb2-1}), we could find the $\delta\bar{l}$ relation as follows
\begin{equation}\label{lb2-3}
  \delta\bar{l}=\frac{2}{k}(l_{t'}-l) \delta\mathcal{P}_0.
\end{equation}

As one can see in the rest of this section, we have found the dictionary between the bulk and the boundary charges in (\ref{hol03}) that can be used to change the $\delta\bar{l}$ relation as follows
\begin{equation}\label{lb2-4}
  \delta\bar{l}=\frac{2}{k}(l_{t'}-l) \delta\tilde{P}_0,
\end{equation}
where the $\delta\tilde{P}_0$ and $\delta P_0^{cyl}$ have a relation as in (\ref{def003}) that can be utilized to find the following relation for the $\delta\bar{l}$
\begin{equation}\label{lb04}
\delta\bar{l}=\frac{4}{k^2}(l_{t'}-l)(P_0^{cyl}-\alpha k)\delta P_0^{cyl}.
\end{equation}

So our conjectural relation for $\delta\bar{l}$ in (\ref{lb01}) is confirmed, and the explicit form of the $f(l, P_0^{vac}, P_0^{cyl})$ can be found as follows
\begin{eqnarray}\label{lb05}
  f(l, P_0^{vac}, P_0^{cyl}) &=& \frac{4}{k^2}(l_{t'}-l)(P_0^{cyl}-\alpha k)\nonumber\\
            &=& \frac{4\alpha^2}{(P_0^{vac})^2}(l_{t'}-l)(P_0^{cyl}-P_0^{vac}).
\end{eqnarray}
where we have used (\ref{Pvac.1}) into the second line of the above relation.

We assume the first law of the entanglement entropy to be true in the gravity side of the $WAdS_3/WCFT$ in the GMMG case (as in the Enstein gravity case in \cite{So1,So2}). Modular Hamiltonian is the same as gravitational charge (\ref{Q04}) , and since we have assumed the first law of entropy to be true, we can write the relation between the holographic entanglement entropy variation $\delta S_{HEE}$ and the gravitational charge variation as follows \footnote {Here we should mention that it is interesting task to one try to apply the proposal and formula (2.29) in the paper \cite{So2} to obtain holographic entanglement entropy in the framework of GMMG model}.

\begin{eqnarray}\label{HS01}
  \delta S_{HEE} &=& \delta Q.
\end{eqnarray}

Now we can complete the dictionaries between the fields in two sides of the duality. The algebra (\ref{VKM03}) of the charges that generate the isometries of the warped $AdS_3$ black hole in GMMG has an equal form with the charge algebra (\ref{VKM02}) in WCFT. This correspondence between the bulk and the boundary symmetry is an indication of the $WAdS_3/WCFT_2$ holography. We could find the following dictionary between the bulk and the boundary variables,
\begin{eqnarray}\label{hol01}
% \nonumber % Remove numbering (before each equation)
  -\frac{1}{2\pi}\tilde{P} &=& -\frac{|\alpha|}{96\pi}c_U M(\phi) \\
  \frac{1}{2\pi}\tilde{T} &=& -\frac{\alpha^4\nu^4}{768\pi}c_V J(\phi). \label{hol02}
\end{eqnarray}

Using these matching relations between the bulk and the boundary variables, we find the following correspondence between the charges in two sides of the duality
\begin{eqnarray}\label{hol03}
% \nonumber % Remove numbering (before each equation)
  \mathcal{P}_m &\to& \tilde{P}_m \\
  \mathcal{L}_m &\to& \tilde{L}_m . \label{hol04}
\end{eqnarray}

Substituting the dictionary variables (\ref{hol01}) and (\ref{hol02}) into the gravitational charge (\ref{Q04}), we find the following relation
\begin{equation}\label{law02}
  \delta Q=\delta\langle \mathcal{H}_{\zeta}\rangle,
\end{equation}
that it is another evidence for the duality is considered in this paper. Using the dictionary (\ref{hol01}) and (\ref{hol02}), the equality between the entanglement entropy (\ref{SR05}) in WCFT and the holographic entanglement entropy (\ref{HS01}) is also established easily. From the equality, $S_{EE}=S_{HEE}$, and using (\ref{law01}) and (\ref{law02}), we implement our claim on the establishment of the first law of the entanglement entropy in the $WAdS_3/WCFT_2$ duality that its final form is as follows
\begin{equation}\label{law03}
  \delta S_{EE}=\delta\mathcal{H}_{\zeta}=\delta Q=\delta S_{HEE}.
\end{equation}

So we have provided a holographic computation of the first law of entanglement entropy for WAdS under the boundary condition (\ref{bC}). Our bulk results in this section agree with those of WCFT results in section \ref{sec:3}, and this is another evidence of the conjectured WAdS/WCFT correspondence.

\section{Conclusion  }\label{sec:6}
In this work, we studied one class of the non-AdS holographies. We studied the entanglement entropy and 1st law in the $WAdS_3/WCFT$ correspondence in the context of Generalized Minimal Massive Gravity. This generalizes similar studies of $WAdS_3/WCFT$ in the Einstein gravity.
 Previously we have studied the $Flat_3/BMSFT_2$  holography in the framework of GMMG model \cite{Setare3}. The field theory at the boundary was considered a theory with $BMS_{3}$-symmetry \cite{Bo,Sa1,Sa2}. The asymptotic symmetry of the bulk side of the duality was the same as the vacuum symmetry of the field theory side. By computing the gravitational charge in the bulk and the modular Hamiltonian at the boundary, we succeeded to show the first law of the entanglement entropy. In this new work, we considered another class of the duality where the filed theory at the boundary is a warped $CFT_2$ \cite{Det1,Hofman1}. In the bulk, the spacelike warped $AdS_3$ black hole \cite{Setare1} as a solution of the GMMG was considered.

The charge algebra of the $WCFT_2$ is a Virasoro-Kac-Moody algebra and this is the same as the asymptotic charge algebra of the bulk at the boundary limit \cite{Setare1}. This is a permission to consider the duality as a true one. We used the modular invariance of the WCFT to compute the modular flow generator at the boundary's vacuum. This was computed using the generalized Rindler transformaton \cite{Casini1,Jiang}. The modular flow generator was computed in (\ref{zeta-2}). To find an evidence for the holography, we needed the modular Hamiltonian and the entanglement entropy in the field theory side. Because of the universal form of the entropy of the WCFT in an interval, we used the result of \cite{Castro}. We expanded this result in details to see the importance of the warped symmetry in calculation of the entanglement entropy.

We showed the first law of the entanglement entropy (\ref{law01}) at the boundary that states the equality of the entanglement entropy and the modular Hamiltonian variations. This form of the first law can be implemented by some variations in the bulk. For the bulk, we used the metric (\ref{Met01}) that is an asymptotically warped $AdS_3$ black hole with the $r$-dependent function (\ref{R01}), (\ref{N01}) and (\ref{Np01}). We found the gravitational charge (\ref{Q01}) with (\ref{gP01}) and (\ref{gL01}).  To establish the first law of the entanglement entropy, introducing the definitions (\ref{M,J}), we found the gravitational charge variations in (\ref{Q04}). This was equal to the holographic entanglement entropy in (\ref{HS01}).
Finding the relations (\ref{hol01}) and (\ref{hol02}) between the variables in two sides of the duality, as a holographic dictionary, we found the completed form of the first law of the entanglement entropy (\ref{law03}).

\section{Acknowledgments}
We thank the anonymous referee for important comments.

\end{document}